# In The Wild Residual Data Research and Privacy

William Bradley Glisson
University of South Alabama
School of Computing Science
Mobile, AL 36688
bglisson@southalabama.edu

Tim Storer
University of Glasgow
School of Computing Science
Glasgow, Scotland
timothy.storer@glasgow.ac.uk

Andrew Blyth
University of South Wales
Faculty of Advanced Technology
Pontypridd, Wales
andrew.blyth@southwales.ac.uk

George Grispos
Lero - The Irish Software Research Centre
University of Limerick
Limerick, Ireland
george.grispos@lero.ie

Matt Campbell
University of South Alabama
School of Computing Science
Mobile, AL 36688
mattcampbell@southalabama.edu

**ABSTRACT**[1]

As the world becomes increasingly dependent on technology, researchers in both industry and academia endeavor to understand how technology is used, the impact it has on everyday life, the artifact life-cycle and overall integrations of digital information. In doing so, researchers are increasingly gathering 'real-world' or 'in-the-wild' residual data, obtained from a variety of sources, without the explicit consent of the original owners. This data gathering raises significant concerns regarding privacy, ethics and legislation, as well as practical considerations concerning investigator training, data storage, overall security and data disposal. This research surveys recent studies of residual data gathered in-the-wild and analyzes the challenges that were confronted. Amalgamating these insights, the research presents a compendium of practices for addressing the issues that can arise in-the-wild when conducting residual data research. The practices identified in this research can be used to critique current projects and assess the feasibility of proposed future research.

---

## 1. INTRODUCTION

Information and communication technologies (ICTs) are now essential parts of the infrastructure of the modern global society. These technologies are becoming ubiquitous, remote and mobile. They enable a multitude of services and information to be delivered to an end-user on demand and on a diverse range of end-user devices. Examples include access to personal information and services, such as social media sites, shopping, banking; social activities, such as online activism and voting; and corporate information, such as accounting systems or business plans. Indicative of this phenomenon, Gartner (2016), for example, has estimated that the combined worldwide shipments of tablets, mobile phones and personal computers (PC) will reach 2.4 billion units in 2016.

Separately, the International Telecommunication Union reported that there were more than 7 billion mobile-cellular subscriptions in 2015 (International Telecommunication Union, 2015). Gartner (2012) has also predicted that annual individual household broadband consumption will increase from 464 gigabytes in 2011 to 3.3 terabytes in 2016.

In order to understand the extent and practice of everyday activities mediated through ICTs, researchers have increasingly turned to a range of 'in-the-wild' methods for data gathering (Hagen, Robertson, Kan, & Sadler, 2005). Chamberlain, et al. (2012) have argued that HCI researchers are increasingly leaving the laboratory and turning towards 'in-the-wild' studies as a better way of understanding user behavior. Separately, Rogers et al. (2011) argues that research conducted in-the-wild can generate results and insights that cannot be obtained in laboratory environments.

The increasing capabilities of digital storage devices has also meant that there is an increasing amount of information left resident long after its usefulness to its original creator has ceased. Recent research by digital forensic researchers has recovered a wealth of personal, sensitive and corporate information from relatively small samples of storage devices. Glisson, et al. (2011) recovered over 11,000 data artifacts from just 49 mobile devices acquired from secondary markets. In additional research, Glisson and Storer (2013) recovered over 7,000 artifacts from 32 mobile phones issued to employees in a Fortune 500 company.

Similarly, studies by Jones (Jones, Dardick, Davies, Sutherland, & Valli, 2009; Jones, Valli, Sutherland, & Thomas, 2008) examined a sample of used hard disks, and reported that over 70% contained personally identifiable information. A similar study conducted by the Real Data Corpus (RDC) project obtained 158 re-sold hard disks, from which they were able to extract 75 gigabytes of data (S. L. Garfinkel & Shelat, 2003).

This residual data is of considerable interest to both researchers in industry and academia for several reasons, for example:

- Residual data research provides insights into the magnitude of privacy and security risks taken by owners of digital devices when they dispose of them. By conducting surveys of residual data on devices sourced from secondary markets, researchers can monitor trends in personal, sensitive and corporate information. This data can be examined to identify risks that can be used to implement successful security breaches. In turn, researchers can evaluate the effectiveness of new security and privacy counter measures, such as custom operating system modifications, in an effort to improve data deletion mechanisms and real-time data breach notification solutions.
- Collection of data sets from residual in-the-wild sources provides digital forensics researchers with realistic test cases on which to evaluate new and existing software tools and practices. Generating realistic data sets artificially, for evaluation purposes, can be extremely difficult due to the variability and idiosyncrasies of individual behavior. Taking data sets from in-the-wild sources provides the opportunity to evaluate forensic tools against targets that can be more representative of real-world investigations. Insights into digital forensic tool performance, potentially, impacts residual data credibility in legal settings. Recent research highlights the importance of residual data

- credibility by indicating that the introduction of residual data as evidence in legal atmospheres is escalating (Berman, Glisson, & Glisson, 2015; McMillan, Glisson, & Bromby, 2013).
- More widely, in-the-wild residual data is of interest to researchers who seek to better understand how users behave with ICTs outside of controlled laboratory settings. The residual data contained on devices, such as message conversations, pictures, application logs and office documents, can be used to reconstruct details of user behavior. This source of information can provide direct insights into how ICTs are used In-The-Wild, rather than how users say they are used. It also initiates discussions on ethical handling of residual digital data.

The investigation of residual data sources raises serious legal and ethical considerations. The very act of investigating the extent of privacy risks through analysis of residual data sources creates the potential for those very risks to be realized, inadvertently or otherwise, by the researcher. Unlike other forms of in-the-wild research, it is unlikely that the former owners of the devices examined will have provided explicit consent for their data to be included in the study which is a standard practice in research involving human participants (British Psychological Society, 2010). Any subsequent analysis of the data found on the storage devices, therefore, raises concerns about the rights and expectations of the device's former owners. In particular, there may be concerns as to their rights or expectations of privacy, which may persist well after the device has been discarded or sold.

Expectations of privacy by the former owners of ICT devices may be complex, given that the definition and understanding of privacy itself is multi-faceted and remains a contested topic (Smith, Dinev, & Xu, 2011). Several authors have conducted literature reviews in the area of privacy, demonstrating the breadth of issues (Belanger & Crossler, 2011; Smith & Milberg, 1996). Despite this complexity and disagreement, it seems clear from the literature that the use of residual data in research raises privacy concerns. Clarke (1999) defines privacy as "the interest that individuals have in sustaining a personal space, free from interference by other people and organizations". The work identifies four aspects of privacy: person, personal behavior, personal communication and personal data (Clarke, 1999); all of which may be affected by the collection of residual data. In addition, Belanger and Crossler (2011) note, in their survey of privacy research, that control over secondary uses of personal information through "the practice of using data for purposes other than those for which they were originally collected" recurs in several definitions of privacy. Therefore, it is essential that studies utilizing in-the-wild residual data be designed to acknowledge and address threats to individual privacy.

This paper contributes to the high-level privacy agenda by reviewing existing research studies of residual data to identify themes. These themes are then used to identify best practices in residual data research and present a compendium of practices for addressing common legal and ethical considerations.

The paper is structured as follows. Section two reviews existing studies of residual data. Section three discusses the ethical and legal challenges that are present in this form of research and describes how the previous studies addressed these challenges. Section four synthesizes this discussion into a compendium of practices for managing in-the-wild residual data during research. Finally, section five draws conclusions and presents future work that should be conducted in this area.

## 2. REVIEW OF RESIDUAL STUDIES

This section reviews the methodologies of previous studies, which recovered and analyzed in-the-wild residual data. The studies were selected due to their visibility in academic literature. The focus and results of the studies discussed in this section are summarized in Table 1. For reader clarification, the investigator referenced in the majority of the paper is the academic primary investigator of a research study, unless otherwise stated.

## 2.1 Device Types and Acquisition Processes

The Garfinkel and Shelat (2003) study describes the collation of data that eventually became the Real Data Corpus, a collection of storage device images used for evaluation of digital forensic software. The Jones, et al. (Jones, Dardick, et al., 2009) study is part of a series of analyses of trends in residual data on hard disks, beginning in 2005. The Glisson, et al. (2011), Jones et al. (2008), and Glisson and Storer (2013) studies investigated residual data on mobile and handheld devices.

Finally, the Jones et al. (2009) and the Szewczyk and Sansurooah (2011) studies investigated residual data on flash storage devices which consisted of external USB and memory cards, respectively. All of the studies broadly followed the same series of activities.

These activities include acquiring their target storage devices, recovering data and analyzing the contents of the data recovered. However, the amount of information provided on an implemented methodology varied from study to study. Both the Glisson, et al. (2011) and the Glisson and Storer (2013) studies provide an extensive description of the processes followed and tools used throughout their experiment. Garfinkel and Shelat (2003) described their recovery process in detail, but also use a range of custom software tools for analysis that are not described in detail.

Table 1: Studies In-The-Wild

| Study | Device Type | | | | | | Sources | | | | Aim | | Results | | | | |
|---|---|---|---|---|---|---|---|---|---|---|---|---|---|---|---|---|---|
| | Hard disk | Mobile | Handheld | USB Flash | Memory card | Reseller | Online auction | Computer auction | Pawnbroker | Computer fair | Privacy risks | Evaluation | Sample size | Identifying | Sensitive | Sanitized | User deleted |
| Szewczyk & Sansurooah 2011 | | | | | ✓ | | ✓ | | | | ✓ | | 119 | 18[b] | | 14 | 89 |
| Glisson, et al. 2011 | | ✓ | | | | | ✓ | | ✓ | | ✓ | ✓ | 49 | 21[c] | 23 | 0 | 15 |
| Jones, et al. 2009a | ✓ | | | | | | ✓ | ✓ | | ✓ | ✓ | | 338 | 55 | | 61 | 83 |
| Jones, et al. 2009b | | | | ✓ | | | ✓ | ✓ | | ✓ | ✓ | | 43 | 23 | | 2 | 20 |
| Jones, et al. 2008 | | ✓ | ✓ | | | ✓ | ✓ | | | | ✓ | | 161 | 10 | | 27 | |
| Garfinkel & Shelat 2003 | ✓ | | | | | ✓ | ✓ | | | | ✓ | ✓ | 158 | | | 12 | 117[d] |

---

[b] Szewczyk and Sansurooah don't report devices containing personally identifying information. The figure shown is the number of devices reported to contain resumes and is taken as a lower bound.

[c] Glisson, et al. did not attempt to determine whether a former owner of a device could be identified through analysis of residual data. The figure shown is the number of devices reported to contain contact database entries and is taken as a lower bound.

[d] Garfinkel and Shelat state that 'practically all' hard disks that were successfully imaged and had not been sanitized contained user deleted files. The figure shown is the total number of hard disks in the study, minus unreadable and sanitized disks.

The sampling methodology is reported in varying detail in the studies. Glisson, et al. (2011) described the procedures followed for selecting devices for purchase from the online auction site (eBay) and pawnbroker sources which included location, pricing and auction format. Jones et al. (Jones, Valli, et al., 2009) and Garfinkel and Shelat (2003) both describe tactics used for acquiring devices. These tactics included buying devices in bulk from major resellers and buying in small lots to avoid raising individual seller's suspicions. Uniquely, Jones, et al. (2009) sampled devices from a number of different countries and used several different research teams to conduct the analysis.

A number of studies by Jones et al. (Jones, Dardick, et al., 2009; Jones, Valli, et al., 2009; Jones et al., 2008) provide an outline of their experimental method and tools used, but the steps taken are not described in detail. Szewczyk and Sansurooah (2011) reported the use of processes and tools similar to that of two of the Jones' studies (Jones, Dardick, et al., 2009; Jones, Valli, et al., 2009).

All but one of the studies reported here acquired mobile devices from a range of secondary markets, including commercial resellers of ICT equipment; online auction sites, such as eBay; public auctions and pawnbrokers. Consequently, the extent to which an intermediary may have altered a storage device before being accessed by the researchers may vary between studies. In some of the studies (S. L. Garfinkel & Shelat, 2003; Glisson et al., 2011; Jones, Valli, et al., 2009), the researchers acquired the devices themselves. However, Jones et al.(Jones, Dardick, et al., 2009; Jones et al., 2008) were supplied with the devices by an intermediary, who kept the original source of the devices confidential while the work was conducted

## 2.2 Protective Measures

Glisson, et al.,(2011) outlined the training procedures they followed with their investigators before allowing them to process the target devices. The investigators were instructed in their legal and ethical responsibilities while working on the study. In particular, each investigator was asked to sign a non-disclosure agreement asserting that they would not discuss the material they viewed within the laboratory with anyone outside of the study team. The investigators were then trained on the software recovery tools using a sample of 'safe' devices known to not contain personal or sensitive information. Details of investigator training were not provided in the other studies.

Glisson, et al.,(2011) describe the laboratory setup and security procedures used to conduct their study. All of the equipment and software used to recover data was housed in a secured and network isolated room. All devices studied during the research were also stored in this location. The room was secured with a fresh lock limiting access to only the researchers and investigators. The equipment, used to analyze the mobile devices, was connected to a local area network within the laboratory. An Internet connected machine located outside of the laboratory was used to assist the research where necessary for tasks like using publicly available databases to acquire additional information on mobile device models. Finally, no personal digital devices of any type were allowed in the laboratory at any time.

Recovered data was temporarily stored on a single machine until its contents had been analyzed. Once checked, data was transferred to a single machine via a local network within the laboratory and backed up to a separate external hard disk. This procedure minimized the risk of illegal data being found on the storage machine or backup and, therefore, avoid having to supply the entire data set to the police. Details of precautions taken in a laboratory setup were not described in the other studies.

In the Glisson and Storer (2013) study, research focuses on residual data from thirty-two mobile devices that were issued to and returned by employees in a global Fortune 500 company. A similar detailed data extraction process was followed that was derived from their previous case study (Glisson et al., 2011).

## 2.3 Profile of the Data Recovered

The exact purpose and method of analysis varied between the different studies. Glisson et al. (2011) categorized each data artefact recovered according to whether it was considered 'personal' if the item could be associated with personal, non-corporate use of the device, 'sensitive' if the item was considered to be potentially compromising for the device's former owner; and/or 'deleted' if the analysis indicated that the user had attempted to remove the item from the device. Garfinkel and Shelat (2003) were concerned with investigating the efficacy of device sanitization practices but also provided commentary on examples of files recovered from the hard disks investigated. All of the Jones studies (Jones, Dardick, et al., 2009; Jones, Valli, et al., 2009; Jones et al., 2008) and the Szewczyk and Sansurooah (2011) study categorized devices in terms of the presence of identifying information, as well as providing informal commentary on files recovered.

All of the studies used the data recovered to investigate the extent of risks to privacy of the device's former owners. In addition, the data gathered by Garfinkel and Shelat (2003) and Glisson and Storer (2013) have been used as the basis for evaluating and/or comparing the performance of digital forensic tools on real-world data sets.

All of the studies reported significant amounts of personal, sensitive, personally identifying and/or corporate data recovered from the devices examined. Table 1, Studies In-The-Wild, summarizes the results presented in the different papers, which are discussed in more detail below.

*Identifying Information*

Jones et al. (2009) reported the number of devices examined which contained personally identifying information broken down by region. In total, the study reported that 62 of the 338 hard disks examined contained information allowing the organization that had previously owned the hard disk to be identified. Also, 55 contained information allowing individuals to be identified. Glisson et al. (2011) did not attempt to determine whether information on devices examined was personally identifying. However, they did recover some 1740 contact database entries from the devices they examined.

*Personally Sensitive*

All of the studies recovered information that could be judged personally sensitive information to the device's former owner. For example:

- Glisson et al. (2011) and Glisson and Storer (2013) reported the recovery of personal photographs containing (adult) nudity, probably taken with the device's camera.
- Garfinkel and Shelat (2003), Jones et al. (2009), and Glisson et al. (2011) all report the discovery of health related information, including medical records, personal conversations about health issues (via SMS) and insurance documentation.
- Glisson et al. (2011) recovered several probable personal identifying numbers (PINs) used in banking disguised as contact information and Garfinkel and Shelat (2003) recovered thousands of credit card numbers.
- Garfinkel and Shelat (2003), Jones et al. (2009), and Glisson et al. (2011) all reported discovering pornographic content on the devices examined. In particular, Jones et al. (2009) reported discovery of pornography on hard disks that were identifiable with commercial organizations.

All of the studies reported finding personal photographs (including family photographs) on devices in their corpus.

*Commercial and Corporate*

The studies of hard disks, USB devices and memory cards reported significant amounts of information from commercial and other corporate contexts. Szewczyk and Sansurooah (2011) reported discovery of business plans, resumes, contact information and pay slips. Jones et al. (2009) found intellectual property material relating to a motor vehicle company, as well as a substantial amount of material from a legal firm. In addition, Jones et al. (2009) also found correspondence concerning currency transactions worth $50 billion.

*Government, Political and Public*

The studies by Jones et al., (Jones, Dardick, et al., 2009; Jones, Valli, et al., 2009) and Szewczyk and Sansurooah (2011) all found information relating to government activities, including documentation from an embassy with minutes of internal meetings; details of a government department's budgets and future plans; and photographs and information on Ministry of Defence (UK) police officers.

Jones et al. (2009) found extensive material from educational organizations, such as schools and universities. These included research material gathered by a university employee and lists of students attending a primary (elementary) school in the United Kingdom.

*Illegal Content*

Jones et al. (2009) reported that 19 of the hard disks examined contained 'illicit' material. Illegally obtained software license keys are reported as an example of this category. In contrast, Glisson et al. (2011) explicitly stated that no illegal content was recovered during their study, despite, for example, discovering SMS conversations discussing illegal drug use.

For all of the studies examined, it is not clear what guidelines or standards were followed in determining that this content was illegal, illicit or otherwise, in the reported studies. It may be, for example, that material was judged to be illicit if it was not actually criminally sanctionable, but could be the subject of a disciplinary investigation in a corporate environment.

*'User Deleted' Content*

Several of the studies reported the extent of 'user deleted' content that was retrieved from studied devices. This is content that the user intended to remove from the device but did not do so successfully. Usually, the presence of the content is no longer presented on the user interface but is still resident on the storage device. This may be due to content de-registration in a file system, or content that has been marked for deletion, but the actual step of removing the bytes of data has not taken place.

Glisson et al. (2011) found some user data on all of the mobile devices examined, however, none of the devices had been completely or successfully wiped. Hence, user deleted artifacts were on 15 (31%) of the devices examined. In contrast, in the Jones et al. (2008) study 27 (17%) of the devices examined yielded no user data. No report is made on evidence of user-deleted data in the study. This difference between the two studies may be accounted for by the differences in the devices examined. A substantial proportion of the devices examined by Jones et al. (2008) appear to be older first and second generation devices, whereas Glisson et al. (2011) focused on newer devices. In addition, from the methodological details available, it appears that Glisson et al. (2011) went to greater lengths to recover data from the devices they examined than Jones et al. (2008).

Glisson et al. (2011) reported that they recovered some 1,934 artifacts as being user deleted, representing almost 17% of their total corpus. Glisson et al. (2011) attempted to investigate the correlation between artefact category (personal, sensitive) and whether the user had attempted to delete them. However, the sample size was too small to make an assessment.

In contrast to the Glisson et al. (2011) study, Garfinkel and Shelat (2003) found that only 12 (9%) of the hard disks they examined had been thoroughly sanitized so that no residual data remained. Of the un-sanitized disks, they report that 'practically' all contained user deleted data. Similarly, Jones et al. (2009) reported that 61 (18%) of the hard disks examined in their study had been thoroughly sanitized. However, in contrast to the Garfinkel and Shelat (2003) study, only 83 (25%) of the disks examined were reported to show evidence of user deletion. This discrepancy may be due to Jones et al. (2009) using a more restrictive definition of 'user deletion' in situations where it is clear that the former owner intended to fully wipe the hard disk, but failed to do so.

Of the two flash storage studies, Jones et al. (2009) reported that only two (5%) of the USB devices studied had been thoroughly wiped but, in contrast, the Szewczyk and Sansurooah (2011) study found that 14 (12%) of the memory cards examined, a far higher percentage, had been wiped. Additionally, 20 (47%) of the USB devices contained user deleted, but recoverable, data as compared with 89 (75%) of memory cards.

*Summary*

Despite the differences in these studies, there is a coherent picture from all of them. A substantial amount of personal and/or sensitive information was found as residual data on a variety of storage device types available from secondary markets. Further, a large corpus of residual data can be recovered from relatively small device samples. The recovery of this residual data presents ethical and legal challenges when conducting studies in-the-wild.

## 3. CHALLENGES IN INVESTIGATING RESIDUAL DATA

This section discusses the legal and ethical challenges that may be encountered when conducting residual data research. Where available, commentary on the issues identified in the other studies is also referenced. Where legal concerns are addressed, the discussion focuses on United Kingdom, European Union and United States contexts, where a large portion, but not all of the work described, was conducted.

### 3.1 Participant Consent

A key principle of studies involving human participants is the need to establish and document the participant's informed consent. In the United Kingdom, exemptions in the Data Protection Act (UK Parliament, 1998) permit legitimate researchers, in some circumstances, to record personal data without informed consent. However, although collection of residual data may be lawful, it may not, necessarily, be ethical. Some of the authors of this paper, for example, are required by their employer to follow the British Psychological Society's (2010) Code of Conduct for research involving human participants.

Typically, participants in a study are given a consent form that describes the purpose of the study, what data will be collected and how it and any derived results will be disseminated. Further, the consent form will normally describe how personal data will be secured and destroyed once the research is complete. In addition, the British Psychological Society's (2010) Code of Conduct requires participant's to be informed, before an experiment begins, that they can withdraw from a research study at any time and any data collected, up to that point, will be destroyed.

Depending on how a device that potentially contains residual data is acquired, these procedures may not be possible for several reasons:

- It may not be possible to identify and/or trace the device's owner before data is recovered and analyzed, particularly if a device has had many owners prior to be being acquired for the purpose of the research. If each previous owner is treated as a study participant, each would need to be informed about the study and prior consent, for the use of their data obtained, before the research could proceed. This may be impossible without accessing data on the device in order to identify previous owners.
- It may be difficult to determine who the 'participants' are, in a study of residual data, for the purpose of acquiring informed consent. For example, a device may have had users who were not the device's owners. In any case, it is not clear that the previous owners or users of a device, or the on-going owners of data on the device, are 'participants' in the accepted sense of the term, since they are not actively participating in a study. However, data about them is being used in the study.
- Rejecting devices for which researchers are unable to obtain informed consent from the original owners and users who generated residual digital data may present an unacceptable study bias. This may be a particular problem where research investigates the extent of personal or sensitive artifacts in a residual data corpus. Former users or owners may be more likely to withhold consent where they believe sensitive or personal data may be found on a device, regardless of assurances about privacy and security from investigators.

Consequently, requiring researchers to obtain informed consent can present insurmountable challenges to studies employing residual data.

### 3.2 Device and Data Ownership

As suggested in the challenges section, there is a distinction between a data owner and a data storage device owner, or data holder. This distinction is codified in the United Kingdom in the Data Protection Act (UK Parliament, 1998).

When an item is purchased, various pieces of legislation impact product and/or service interaction. Legislation relating to contract law is the primary piece of legislation governing this area. However, consumers are accustomed to treating ICT equipment, which is purchased every day, as physical artifacts. Hence, end-users are lured into treating the contracts governing the disposal of ICT equipment and data storage devices in the same manner. This creates an environment where an organization sells all of its old equipment at the end of its life to a reseller and the reseller collects physical artifacts and considers that to be the end of the matter.

This view is, however, incorrect because under the UK Data Protection Act an ICT data storage device is viewed as not only a physical device but, also, a logical device. The logical device has a unique set of legal requirements that cannot be transferred. The Logical device will contain not only personal and corporate data, but also software and operating system licenses.

There are many different types of software licenses that can fall into the realm of software license management. These license types range from single user license, to original equipment manufacturer license bundles, to education software agreements. Some licenses state that the license is specific to the machine on which the software is executing and that the license to use the software becomes null and void when the machine is recycled and destroyed. Others provide the ability for a user to un-register a machine and to re-register the software to another machine. While research investigates the cost of software re-use, the actual figures on how many organizations fail to take advantage of this functionality and, thus, incur the cost of the lost license is not known. From a software piracy perspective, the ability to harvest and resell software licenses from second hand computer hard-drives offers the ability to potentially generate significant income.

Under the data protection act, the organization that generates the personal data is responsible for processing the data. The data protection act defines a data controller as "...a person who (either alone or jointly or in common with other persons) determines the purposes for which and the manner in which any personal data are, or are to be, processed" (UK Parliament, 1998).

According to the definition in the data protection act, a data controller or data controllers must decide the purposes for which and the way in which personal data are, or will be, processed. The determination of the purposes for which personal data are to be processed is paramount in deciding whether or not an individual or organization is a data controller. When a person determines the purposes for which personal data is to be processed, a decision as to the manner in which those data are to be processed is often inherent in that decision. In particular:

- Personal data shall be obtained only for one or more specified and lawful purposes, and shall not be further processed in any manner incompatible with that purpose or those purposes.
- Appropriate technical and organizational measures shall be taken against unauthorized or unlawful processing of personal data and against accidental loss or destruction of, or damage to, personal data.
- Where processing of personal data is carried out by a data processor on behalf of a data controller, the data controller must, in order to comply with the seventh principle: choose a data processor providing sufficient guarantees in respect of the technical and organizational security measures governing the processing to be carried out, and take reasonable steps to ensure compliance with those measures.

Therefore, the act places certain responsibilities on data holders regarding the custody of data on behalf of data owners. European legislation provides for similar regulations. The implication for studies of residual data in the United Kingdom is that former owners of data storage devices continue to retain rights over the data on the device even after they have sold them. Once the researchers have obtained the device and accessed the data, they are acting as data holders with a responsibility to ensure the integrity and security of the data on behalf of the data owners. These responsibilities exist even though the former owners of the device will be unaware of the use of the data.

A consequence of these responsibilities is apparent in the way that results are disseminated from residual data studies in the United Kingdom. All of the studies reviewed that were based in the United Kingdom provided only abstracted information on the types of data artifacts found on the devices studied. In addition, none of these studies (as far as is known) have published the raw data on which their reports are based. In contrast, the data gathered by the (S. L. Garfinkel & Shelat, 2003) study which was based in the United States has been compiled into a corpus of storage device images that are made available to legitimate researchers (S. Garfinkel, Farrell, Roussev, & Dinolt, 2009).

A final aspect of data ownership concerns disposal. None of the studies describe how devices were disposed of following the experiment. Indeed, in the (S. L. Garfinkel & Shelat, 2003) study, the data recovered is now available as a corpus for evaluating forensic software tools and will, hopefully, be maintained by the researchers indefinitely.

In the United Kingdom, researchers are under a legal and ethical obligation to ensure that data collected in a research study is destroyed once the work is complete (British Psychological Society, 2010; UK Parliament, 1998). Researchers must ensure that storage devices that are used for data analysis and the original devices containing the residual data are effectively erased. This can be particularly challenging to achieve when researchers are working with certain categories of storage devices such as embedded flash chips. This is due to less than effective deletion facilities on individual devices.

### 3.3 Data Protection Laws

Many countries have laws designed to protect data privacy and security. For instance, accessing data on a digital storage device without the permission of the content's owner(s) could be considered a violation of the Computer Misuse Act (UK Parliament, 1990) since it is unlikely that the former owner of a device conceived or consciously gave permission for residual data to be accessed by researchers.

A specific case concerns access to content that is protected by encryption or access control mechanisms. In this situation, attempts by researchers to circumvent controls could be considered criminal activity. This situation is complicated because many of the software tools (such as digital forensic toolkits) contain features to help investigators bypass controls. In some circumstances these features are applied automatically to a device, or the toolkit is configured to use them by default.

One thing to consider is whether it is necessary for the purposes of the experiment to attempt to circumvent these controls. If the research is only concerned with measuring readily accessible data, for example, then storage devices that have been encrypted may not require further examination.

Even in countries with arguably less stringent data protection laws (such as the US), the Health Insurance Portability and Accountability Act (HIPAA) of 1996 prohibits the transfer of personal health information (PHI) from a covered entity (e.g. healthcare provider, insurance company, or clearing house) to an entity not involved in patient treatment or payment for treatment without the express consent of the individual to which the data pertains. HIPAA does allow a covered entity to share PHI with their business associates provided they agree to provide the same protections as the covered entity (CMS, 2013).

## 3.4 Encountering Illegal Information

There is a possibility that a storage device acquired from secondary sources may contain data that is considered illegal in a given jurisdiction. Examples include digital evidence of conventional criminality such as fraud, drug dealing or even terrorism; and criminal activity more usually associated with ICT, notably illicit pornographic images or videos.

The difficulty for researchers investigating residual data is in ascertaining the contents of a storage device acquired from the secondary markets without viewing illegal material, which could be a criminal act in some jurisdictions. From the research study's perspective, there is the risk that any illegal content will also 'contaminate' legal material recovered from other storage devices, since police investigators may need to confiscate all storage devices identified as containing illegal material.

Glisson et al. (2011) described a procedure for minimizing this risk in their study. The procedure involved storing data recovered from a device on a single PC while it was examined by an investigator. Once it had been determined that the data did not contain any illegal content, it was added to the corpus. This meant that should illegal content be discovered, only the hard disk from the PC used for analysis would need to be supplied to the police.

Glisson et al. (2011) also established procedures for research investigators to report illegal content to legal authorities if and when it was discovered. The research investigators were trained to immediately report the discovery of potentially illegal content, whenever they were in doubt, to one of the lead researchers. If the researcher determined the content was illegal, the hard disk on which the content was stored was to be removed and transferred to a safe until it could be provided to the police. This procedure minimized the risk of researchers or investigators being prosecuted for holding illegal content.

## 3.5 Risk to Stakeholders

From these studies we have identified five unique categories of individuals who may experience risk arising from a study of residual data on ICT devices. These include:

- Previous owners
- Previous device users
- Others who have interacted with previous owners or users
- Researchers and Research assistants
- Current device owners (if not owned by the researchers)

We will discuss the risks to each group.

*Risk to Previous Owners*

A blatant threat to the previous device owner(s) is the risk of privacy violations. For many individuals, ICT devices function as a modern day multimedia diary complete with photos, text and email conversations, and even GPS logs. A secondary, but related, risk involves user security. Risks in this category can involve issues of personal and financial security. Personal security could be compromised if device data can be used to predict where an individual might be at any given time (Grispos, Glisson, Pardue, & Dickson, 2015). Financial security could be compromised if account passwords or near field communication payment information is recovered from the device. Lastly, legal liability can arise if the previous owner(s) engaged in illegal behavior with the device, such as hacking or viewing child pornography. Information recovered from the device could be used as evidence in legal proceedings. Legal liability can also arise when devices are used for legal activities (such as a physician accessing patient records) and practitioners fail to take steps to ensure that residual data does not remain on the device (HIPAA violation).

*Risk to Previous Device Users*

In some cases, the previous users of the device were never the owners of the device. This can occur in situations when an organization issues devices to its employees for work-related purposes. The risk to these users is similar to those of previous device owners, however, privacy expectations are, generally, lower when using corporately-owned devices.

*Risk to Others Who Have Interacted With the Device*

While most studies focus on the risk to those who have actually come into contact with the device, it should not be forgotten that the main function of ICT's is communication. Therefore, those who have communicated with the device being studied are also at risk. Residual text messages, emails, and pictures are good examples of sensitive data that can be left behind on a device and present a risk to an individuals' privacy.

*Risks to researchers and research assistants*

Researchers and research assistants encounter different kinds of risk when conducting research in this area. An obvious risk is the discovery of data that initiates involvement in a legal investigation and an accompanying loss of data. Some of the studies examined earlier in this paper detailed the extensive measures taken to ensure that any illegal data found would be turned over to the authorities and that the illegal data would be isolated from the rest of the data collected so that researchers would not lose access to all of their data if computers, hard drives, and devices were turned over to the authorities. A second, but less obvious, risk is the psychological risk associated with exposure to data of a highly personal nature. Previous studies indicate that there is an elevated risk of depression among those who deal with the sensitive personal data of others such as social workers and those in healthcare (Health.com, 2010). Hence, it is plausible that having access to large amounts of sensitive personal data could trigger depression in researchers as well.

*Risks to device owners (if not owned by the researchers)*

Lastly, organizations or individuals who make devices available to researchers also face some risk depending on the country in which the research is being conducted. As noted earlier, data protection laws in Europe and other places can be more stringent than those in the United States. While in the US, ownership of data transfers with the device on which the data resides, this is not true in some countries (e.g. UK). These laws could open device owners up to legal liability for allowing researchers to access owned devices. Even in countries with less restrictive data protection laws, legal liability can arise from organizational data (such as patient treatment data) being accessed by researchers outside of the organization unless agreements are signed beforehand that protect the organization and its data.

The various risks acknowledged in this section prompted deliberations focusing on protecting stakeholders and mitigating risk. In an attempt to achieve this goal, the following residual data research practices have been compiled.

## 4. RECOMMENDED RESIDUAL DATA RESEARCH PRACTICES

This segment presents recommended practices for conducting research using in-the-wild residual data. These practices are based on the experiences reported in the sections investigating the Review of Residual Data Studies and identifying the Challenges in Investigating Residual Data.

### 4.1 Identifying Research Objectives

Before beginning a research study using residual data it is important to be clear about the objectives and rationale for the work. In particular, because of the sensitivity of material that may be gathered, it is important to ensure that the use of residual data is essential for the objectives of the study. A study to determine the extent of privacy risks on a particular category of storage devices can be justified in using residual data. Research work to develop a realistic corpus of material for evaluating forensic software can also be justified, since there are currently no other methods for constructing such a corpus. In contrast, studies that intend to use devices obtained from the secondary market purely because they are cheaper are unlikely to receive ethical approval. In these circumstances, the justification is not based on the objectives of the research, but on cost grounds, which is normally inadequate.

Similarly, when devising data recovery methods for acquired devices, it is important to consider what tools are appropriate to meet the research objectives and minimize privacy threats to former device owners. For example, a study that is concerned with monitoring trends in sanitization procedures may not need to use tools, which visualize the contents of a storage device. Tools or scripts, which only determine whether a device has been thoroughly sanitized, may be sufficient.

### 4.2 Investigator Training

Providing investigators with the foundation for conducting research on devices that contain real data is critical. Investigators need to be trained on how to use data extraction tools whether they are commercially available or open source.

Training not only impacts the amount of overall data that is potentially recoverable but it also helps to deter contamination possibilities. Investigators need to be made aware that they may be exposed to unpleasant information and even potentially illegal information. Procedures need to be explicitly identified and presented to the investigators on how to handle these situations. This training should also include educating investigators on the importance of data privacy and the signing of non-disclosure agreements.

There is the possibility that investigators of digital devices could run across passwords to email, cloud storage services or social networking web sites. Investigators need to be instructed to never, under any circumstances, access data on outside systems through the device they are examining or other computers involved in the extraction or analysis of residual data. In addition, access to extracted data should be highly regulated.

### 4.3 Device Acquisition Management

Storage devices containing residual data can be obtained from a variety of sources and under different circumstances, for example:

- donated by volunteers;
- purchased from secondary markets, such as pawnbrokers and auction sites;
- loaned from resellers or original equipment manufacturers, prior to destruction;
- loaned from a third party that maintains its own device corpus;
- obtained from lost property departments;
- acquired from a corporate IT department's device life-cycle management processes.

The source of residual data may have implications for data ownership, and consequently for the methods employed in recovering, analyzing and disposing of data, depending on the jurisdiction in which research is conducted.

Issues of device and data ownership, when sourcing from secondary markets, were discussed in the challenges section. Different constraints may arise, for example, if the devices are on loan, since there may be an obligation to return the devices to the provider unaltered. In these situations, it may be necessary for researchers to establish what actions will be taken with a device set after it is returned to the provider and factor these constraints into the study's objectives and methodology. Researchers may be required to transfer possession of storage devices to criminal investigators if illegal content is discovered.

Devices acquired from a corporate IT department may be subject to very different legal and ethical constraints. It is not unusual for corporate organizations to require users of IT equipment to transfer ownership of data created and/or held on the devices to the organization. Consequently, researchers may be under relaxed legal constraints when working with residual data sourced from corporate environments, since they may never act as data holders. However, work in this context may still give rise to ethical dilemmas. For example, results of a study of residual data might reveal examples of misbehavior by employees of an organization. It is not clear whether the results of studies of residual data should be available to managers in an organization in order to take disciplinary action. Regardless of the ethical position, researchers should at least be clear about the objectives of a research study with an organization and the limits to which results can be used, before data recovery and analysis commences.

### 4.4 Data Anonymization

The anonymization of extracted residual data is beneficial to all parties involved, directly or indirectly, with the research. It protects potentially sensitive information of participants like credit card numbers, national insurance numbers, Global Position System (GPS) data on images stored on devices, medical appointments, and sensitive company documents. Hence, all data should be anonymized through the presentation of the information in broad categories and generalities.

### 4.5 Lifecycle Security Practices

A description of security measures for protecting participant data is a standard part of approval and consent procedures for studies involving human participants (UK Parliament, 1998). Details typically include a description of how data will be stored and who will have access to the data. However, security practices with residual data need to be considerably more robust than for data recorded for conventional experiments with human participants. As noted above in the Challenges section, it is unlikely that data owners will have provided explicit consent for data to be gathered. Consequently, any data recovered may contain information that a 'participant' might not ordinarily volunteer. In addition, it is important to develop practices that ensure data integrity should it be necessary, at a later point in time, to transfer this information to the police for a criminal investigation.

Glisson et al. (2011) described their procedures for securing the residual data they recovered from mobile devices using a combination of physical and electronic measures. All equipment and acquired devices were stored in a physically secured and network isolated laboratory. Personal electronic devices not associated with the research were not permitted in the room. The hard disks of the machines used to recover data were erased after the experiment was complete. The corpus was stored on an encrypted and physically secured hard disk for future analysis. Finally, the corpus of devices was stored in a physically secure location after data had been recovered.

As noted above, none of the studies addressed the disposal of residual data and devices after the completion of an experiment. This is, however, a crucial aspect of ensuring the on-going privacy of the device's former owners. Re-selling devices on secondary markets may be a tempting proposition to reduce costs associated with a study. However, storage devices need to be thoroughly erased before decommissioning and disposal. Where it cannot be guaranteed and demonstrated that this has occurred (when investigating embedded storage devices, for example), it may be necessary to contract a reputable organization to ensure that the entire device is destroyed.

A further option to consider is obtaining independent audits of the security procedures employed while residual data is being held for research. This can be undertaken on a reciprocal basis by other researchers with experience in conducting research with residual data.

### 4.6 Ethics Committee Liaison

Research utilizing in-the-wild residual data is still relatively scarce, so it is unlikely that the members of an institution's ethics committee will have previous experience in handling both legal and ethical issues. Hagen et al. (2005) has argued that ethics committees are (understandably) ill-equipped to manage the legal and ethical issues involved in 'in-the-wild' research in general.

Consequently, effective liaison between the research team and the relevant institution's ethics committee is essential throughout the conduct of the study. As well as explaining the usual objectives and rationale for the research, applications for ethical approval for residual data research should be accompanied by:

- a justification for the use of residual data in the research;
- an explanation of why the research cannot be effectively conducted using non-residual data sources;
- a description of the ethical concerns and risks raised by the use of residual data; and
- a description of the steps taken to address or mitigate these concerns.

Contact should be maintained between researchers and the ethics committee during the gathering and analysis of residual data. In particular, it is important for the ethics committee to be informed should any of the risks described during the approval process be manifested. Demonstrating competence in managing these issues can help alleviate concerns as to the risks for future research. However, researchers should accept that in some circumstances it may be necessary for an ethics committee to halt a research study when such risks are manifest.

Finally, it is helpful to provide a summary report to the ethics committee after the research is complete. Supplying ethics committee members with papers describing the results of the research can help ease approval processes in this difficult area for future research studies.

### 4.7 International Residual Data Gathering

Of the studies examined, only the Jones et al. (2009) research appears to have attempted to compare residual data risks between multiple countries or legal jurisdictions. In their case, data recovery and analysis took place in each country (Australia, France, Germany, United Kingdom and United States) independently, with only the summary information from each country collated in the study findings.

Policy makers in the European Union and the United Kingdom have already begun to suggest that data transfer beyond their borders may be problematic, due to differences in regulatory approaches to data protection (Information Commissioner's Office, 2012; Working Party, 2012). Wolthusen (2009) has argued that these regulatory issues will make criminal investigation of digital crimes across borders more problematic in the future. It is likely that these complexities will be no less problematic for privacy researchers.

Regulation of data ownership and rights appear far more permissive in the United States than in the European Union. It is unclear whether, for example, the Data Protection Act (UK Parliament, 1998) would apply to residual data collected in the United States and transferred to the United Kingdom. Certainly, it would seem that regulations prevent a similar transfer of (non-residual) data to the United States (Grispos, Storer, & Glisson, 2012). Any researcher intending to transfer raw residual data across international boundaries should certainly seek legal advice before doing so.

### 5. CONCLUSIONS

This research reviewed the available studies that have used residual data In-The-Wild. The studies to date have focused on residual data recovered from several different digital storage technologies (hard disks and flash media chips) integrated in a range of ICTs, including desktop and laptop personal computers, servers, external USB storage, mobile devices and handheld devices. These early studies have already demonstrated the extent of personal, sensitive and identifying information available from residual data sources. This work has meant that the first steps in estimating the risks of residual data to individual and corporate privacy have begun to be investigated and quantified. The work has also provided impetus for ICT vendors to improve life-cycle protections for user privacy in the products they offer.

Investigating residual data for research purposes carries its own privacy risks, as identified in this paper. The synthesis of the study methodologies presented in this paper provides a compendium of good practices to guide other researchers working with residual data. Future studies of residual data may seek to integrate and adapt these practices in a variety of ways. Researchers in the future, for example, may seek to use sources of residual data to validate findings from other research methodologies.

Alternatively, researchers may begin to address the residual data risks in other ICTs that have not been examined to date. The data sources considered so far concern relatively small-scale storage capacities. However, data storage is increasingly becoming remote and virtualized. Social media applications, for example, enable users to record huge quantities of data about their private lives, which may be difficult to alter or erase if an account becomes abandoned. These environments present both new opportunities and risks for researchers working with residual data. Existing principles and practices will need to be substantially altered, if residual data methodologies are to be used in recovering, analyzing and understanding residual data in these environments.

## 6. ACKNOWLEGMENTS

This work was partially supported by Science Foundation Ireland (SFI) Grants 10/CE/I1855 and 13/RC/2094, and ERC Advanced Grant no. 291652 (ASAP).